\documentclass[twocolumn,showpacs,amsmath,amssymb]{revtex4}

\def\Lie{\pounds}
\def\beq{\begin{equation}} \def\eeq{\end{equation}} 
\def\four{\,{}^{(4)}\kern-1.5pt}

\begin{document}

\title{New Minimal Distortion Shift Gauge}

\author{Robert T. Jantzen}
\email{robert.jantzen@villanova.edu}
 \affiliation{Department of Mathematical Sciences, Villanova University, Villanova, PA 19085 USA}
 \altaffiliation[Also at ]{ICRA, University of Rome, I--00185 Rome, Italy}
\author{James W. York, Jr.}%
 \email{york@astro.cornell.edu}
\affiliation{%
Department of Physics, Cornell University, Ithaca, New York 14853
}%

\date{April 4, 2006}
\begin{abstract}
Based on the recent understanding of the role of the densitized lapse function in Einstein's equations and of the proper way to pose the thin sandwich problem, a slight readjustment of the minimal distortion shift gauge in the $3+1$ approach to the dynamics of general relativity allows this shift vector to serve as the vector potential for the longitudinal part of the extrinsic curvature tensor in the new approach to the initial value problem, thus extending the initial value decomposition of gravitational variables to play a role in the evolution as well. 
The new shift vector globally minimizes the changes in the conformal 3-metric with respect to the spacetime measure rather than the spatial measure on the time coordinate hypersurfaces, as the old shift vector did.
\end{abstract}

\maketitle

\section{Introduction}

In Minkowski spacetime, global Lorentz frames are available in which one can use the standard coordinates $(t,x,y,z)$ or any 3-dimensional coordinate system to replace the Cartesian coordinates $(x,y,z)$. The choice of space coordinates and their continuation in time is therefore simple. The time vector $\partial/\partial t$ is orthogonal to the flat (Euclidean) spatial hyperplanes, and there are no off-diagonal metric components.

In the curved spacetimes encountered in general relativity, the situation is quite different. In constructing a spacetime, one has two problems related to time. First is the choice of a slicing of spacetime by 3-dimensional hypersurfaces $t=\hbox{\it const.}$ with induced metrics $g_{ij}$ ($i,j=1,2,3$) that are uniformly of Euclidean signature on each ``leaf" of the ``foliation" (slicing). This step requires that the scalar function $t$ (units $c=1$) has an exterior derivative $dt$ such that the vector field $\four g^{\mu\nu} (dt)_\nu$ is timelike ($\mu,\nu=0,1,2,3$). This condition does not imply that the vector field $\partial/\partial t$ dual to $dt$ is necessarily timelike. Whether $\partial/\partial t$ is timelike depends on the magnitude of the spatial shift vector field $\beta^i$. See \cite{york79} for a coordinate-free discussion of this matter and a simple diagram. Analytically, $\partial/\partial t$ is inside the light cone if and only if $g_{ij} \beta^i \beta^j < N^2$, where $N>0$ is the lapse function, $N=[-{}\four g^{-1}(dt,dt)]^{-1/2}$. The shift vector field determines how the spatial coordinates on one slice are pulled to the next nearby slice.

In this paper we introduce a new ``minimal distortion" shift vector resting intrinsically on spatial kinematics and dynamics in spacetime. We regard the new shift as an improvement on what was previously called the ``minimal distortion" shift vector, so named since it minimizes the change in the conformal 3-metric, but with respect to the spatial measure on the time coordinate hypersurfaces rather than the spacetime measure.  That vector was not explicitly linked to the dynamics \cite{smarryork2}. 

We will incorporate two viewpoints concerning the initial value problem. In the Cauchy initial value problem, the data on the ``initial" spacelike slice are the induced metric $g_{ij}$ and the extrinsic curvature tensor $K_{ij}$. The latter measures curvature intrinsic to the slice with respect to that of the ambient spacetime in which it is embedded. The extrinsic curvature makes no reference whatsoever to the lapse $N$ nor to the shift $\beta^{i}$. (See the Appendix of \cite{york99} for a detailed discussion.) This is the ``one-surface" form of the initial value problem.
The other view of the initial value problem has been called the ``thin sandwich" or two-surface form of the initial value problem because it refers to two infinitesimally separated, non-intersecting slices. In this form, $K_{ij}$ is replaced by the velocities, using $\partial/\partial t$, of certain metric components. This procedure necessarily brings in explicit dependence on $N$ and $\beta^{i}$ because it makes reference to the spacetime coördinates in the neighborhood of the single initial slice. The consistency of the method of solution of the initial value problem in either of these two viewpoints is discussed explicitly in \cite{york99}.
In view of the fundamental differences in the two viewpoints above, it is significant that the new minimal distortion shift vector plays a role in both.

\section{Old and New Minimal Distortion Shifts}

Starting with a timelike foliation of the spacetime and adapted coordinates $\{x^\alpha\} = \{x^0=t,x^i\}$, the spacetime metric line element
\begin{eqnarray}
  ds^2 &=& \four g_{\alpha\beta} dx^\alpha\, dx^\beta
  \nonumber\\
       &=& -N^2 dt^2 + g_{ij} (dx^i + \beta^i dt)(dx^j + \beta^j dt)
\end{eqnarray} 
can be re-expressed in terms of the lapse function $N$, the shift vector field $\beta^i$ and the spatial metric $g_{ij}$ (inverse denoted by $g^{ij}$), variables which trace back to Choquet-Bruhat \cite{yvonne56}, 
Dirac \cite{dirac} and Arnowitt, Deser and Misner \cite{adm} and which were aptly named by
Wheeler \cite{wheeler}. 
In turn the lapse may be expressed in terms of a new metric variable which has been called by various names including the densitized lapse, the lapse anti-density, the slicing function or simply the Taub function
\beq
  \alpha = N/g^{1/2} \leftrightarrow N = g^{1/2}\alpha\ ,
\eeq
which is an oriented weight $-1$ spatial scalar density, where $g=|\det(g_{ij})|$ and $\four g =|\det(g_{\alpha\beta})|=N^2 g$ are the absolute values of the spatial metric and spacetime metric determinants. The unit normal $n^\alpha$ to the hypersurfaces $\Sigma_t$ of constant coordinate time $t$ is then in index-free form
\beq
n=N^{-1}(\partial/\partial t -  \beta^i \partial/\partial x^i)
\leftrightarrow
\partial/\partial t = N n + \beta^i \partial/\partial x^i
\ .
\eeq
Taub was the first to make significant use of the function $\alpha$ by choosing the natural gauge $\alpha=1$ to find his two famous Bianchi type II and IX solutions of the vacuum Einstein equations \cite{taub51,bobelba,hdcm}.

The extrinsic curvature is a symmetric spatial tensor whose mixed form is convenient to decompose into its pure trace and tracefree parts
\beq
  K{}^i{}_j 
   = -\frac1{2N} (g{}^{ik} \dot g_{kj} 
                 -g{}^{ik}\Lie_\beta g_{kj} )
  = A{}^i{}_j +{\textstyle\frac13} \tau \delta^i{}_j
\ ,
\eeq
where $\dot f = \partial f/\partial t$ is the ordinary partial derivative and $\tau=K^i{}_i$ is the trace. 
This tensor is subject to the supermomentum constraint
\beq
\nabla_j (K{}^j{}_i - K{}^k{}_k \delta^j{}_i) = j_i
\ ,
\eeq
which can be thought of as a condition on the divergence of the tracefree part of the extrinsic curvature, determining it in terms of the trace $\tau$ and the source current
$j_i= -n^\alpha T_{\alpha i}$  \cite{york73}
\beq\label{eq:smcon}
\nabla_j A{}^j{}_i = \frac23 \nabla_i \tau + j_i\ .
\eeq

The original minimal distortion shift gauge \cite{york79,smarryork1,smarryork2} was chosen to minimize the square of the time rate of change of the conformal metric 
$g^{-1/3} g_{ij}$ (a tracefree tensor density of weight $-2/3$ since $g^{1/2}$ has weight 1)
when integrated with respect to the usual spatial volume element measure over a region of a hypersurface and varied with respect to the shift vector field $\beta^i$ with fixed variations on the boundary \cite{smarryork2}. The mixed form of this coordinate time derivative is
\begin{eqnarray}
 \Theta^i{}_j &=&
g^{1/3} g^{ik} (g^{-1/3} g_{kj})\,\dot{}
= g^{ik} (\dot g_{kj}-{\textstyle\frac13} g_{kj} g^{mn}\dot g_{mn})
\nonumber\\
&=& -2N A^i{}_j + [L \beta]^i{}_j\ ,
\end{eqnarray}
where
\begin{eqnarray}
  [L \beta]^i{}_j 
   &=& g^{ik} \pounds_\beta g_{kj} 
         -{\textstyle\frac13}\delta^i{}_j g^{mn}\pounds_\beta g_{mn}
\nonumber\\
   &=& g^{1/3} g^{ik} \pounds_\beta (g^{-1/3} g_{kj})
\nonumber\\
  &=& 2 g^{ik}\nabla_{(k} \beta_{j)} - (2/3)\delta^i{}_j \nabla_k \beta^k 
\end{eqnarray}
is the zero weight rescaling of the
tracefree mixed-index form of the Lie derivative of the conformal metric with respect to the vector field $\beta^i$, the fully covariant-indexed form of which can be thought of as a tangent vector to the orbits of the spatial diffeomorphism group on the space of conformal metrics, and $\nabla_i$ is the spatial covariant derivative. The corresponding action integral is
\beq
{\rm ActionOld}[\beta,t)
 = \int_{\Sigma_t} \Theta^i{}_j \Theta^j{}_i g^{1/2} d^3 x\ ,
\eeq
and its variation is
\begin{eqnarray}
\delta {\rm ActionOld}[\beta,t)
 &=& 2\int_{\Sigma_t} \Theta^i{}_j [L(\delta\beta)]^j{}_i g^{1/2} d^3 x
\nonumber\\
 &=& 4\int_{\Sigma_t} \Theta^i{}_j \nabla_i\delta\beta^j g^{1/2} d^3 x
\nonumber\\
&=& -4\int_{\Sigma_t} \nabla_i \Theta^i{}_j \delta\beta^j g^{1/2} d^3 x\ ,
\end{eqnarray}
provided one can safely ignore the boundary term in the integration by parts, e.g., by assuming that the variation $\delta\beta^i$ vanishes there. The old minimal distortion (``shear") equation is then
\beq
 \nabla_i (-2N A^i{}_j + [L \beta]^i{}_j) =0\ .
\eeq

Motivated by the new understanding of the initial value problem \cite{york99,py03}, which in the thin sandwich approach \cite{york99} utilizes the lapse-corrected time derivative and  lapse-corrected spatial Lie derivative as the two contributions to the extrinsic curvature, we find that it is natural to rescale the quantity inside the divergence by $N^{-1}$, which can be accomplished by instead minimizing the square of the lapse-corrected time derivative of the conformal metric (corresponding to $N^{-1} \partial/\partial t$ instead of just $\partial/\partial t$, reflecting the proper time elapsed along the unit normal to the time coordinate hypersurface), and using the square root of the absolute value of the full spacetime metric determinant in the measure \cite{py03}
\beq
{\rm ActionNew}[\beta,t)
 = \int_{\Sigma_t} (N^{-1}\Theta^i{}_j) (N^{-1}\Theta^j{}_i) \, N g^{1/2} d^3 x\ .
\eeq
Now the same variation leads to
\begin{eqnarray}
&&\delta {\rm ActionNew}[\beta,t)
\nonumber\\
&&\qquad = 2\int_{\Sigma_t} (N^{-1}\Theta^i{}_j) N^{-1}[L(\delta\beta)]^j{}_i \, N g^{1/2} d^3 x\ ,
\nonumber\\
&&\qquad = -2\int_{\Sigma_t} \nabla_i (N^{-1}\Theta^i{}_j) \delta\beta^j g^{1/2} d^3 x\ ,
\end{eqnarray}
so that the new minimal distortion equation is (dividing by an extra factor of 2 as well)
\beq
 \nabla_i (- A^i{}_j + (2N)^{-1} [L \beta]^i{}_j) =0\ .
\eeq
This can be interpreted as saying that $N^{-1}\Theta^i{}_j$ is orthogonal to $[L(\delta\beta)]^i{}_j$ in the original spatial metric measure inner product without the lapse factor, or that the two lapse-corrected quantities are orthogonal in the spacetime measure inner product. 

The only difference between the two equations is whether an overall factor of the lapse is left inside the covariant derivative expression or is pulled outside.
The distinction between the new and old minimal distortion shift equations therefore evaporates when one considers a spacetime and a splitting in which  $N$ is a constant, since then it passes through the derivative and the two equations agree.
This is the case in spatially homogeneous cosmology \cite{bob80}.

\section{Initial value problem: Conformal Considerations}

To understand the significance of this change in the minimal distortion shift equation, the conformal approach to the initial value problem must be discussed. Recent work
points to the Taub function $\alpha$ as the true time gauge variable rather than the lapse function itself, which together with the (contravariant!) shift vector field as a generator of spatial diffeomorphisms should be held fixed under the conformal rescalings of the $3+1$ variables needed to solve the initial value problem.

We adopt the notation of Pfeiffer and York \cite{py03}: let an overbar denote the physical metric variables, which are related by a conformal rescaling to the unphysical variables without an overbar. The Taub function and the shift vector field and the trace $\tau$ of the extrinsic curvature are not transformed
\begin{eqnarray}
\bar\alpha &=& \alpha\ ,\
\bar \beta^i = \beta^i\ ,\
\bar g_{ij} = \phi^4 g_{ij}\ ,
\nonumber\\   
\bar g{}^{1/2} &=& \phi^6 g{}^{1/2}\ ,\  
\bar N = \phi^{6} N\ ,
\end{eqnarray}
so that the lapse function $N=\alpha g^{1/2}$ transforms by the same factor as $g^{1/2}$. 
The fixing of the Taub function under the conformal transformation is motivated by the correct properties of the Einstein equations in phase space that occur when this quantity is fixed in the canonical action principle \cite{andyork98}.

The extrinsic curvature expressed in terms of the metric velocity and shift vector field is
\beq
  \bar K{}^i{}_j 
   = -(2 \bar N)^{-1} (\bar g{}^{ik}\dot{\bar g}_{kj}
                 -\bar g{}^{ik}\Lie_\beta \bar g_{kj} )
   = \bar A^i{}_j +{\textstyle\frac13} \bar K^k{}_k \delta^i{}_j
\ ,
\eeq
so that its tracefree part is
\beq\label{eq:Athinsand}
 \bar A^i{}_j 
= (2 \bar N)^{-1} [ 
   -(\bar g{}^{ik}\dot{\bar g}_{kj} - {\textstyle\frac13} \delta^i{}_j \bar g^{kl} \dot g_{kl})
  +(\bar L \beta)^i{}_j]
\ .
\eeq
The mixed form of the Lie derivative term appearing in the extrinsic curvature
has the transformation law
\beq
  \bar g{}^{ik} \Lie_\beta \bar g_{kj}
     =  g{}^{ik} \Lie_\beta  g_{kj} + 4 \ln\phi_{,k} \beta^k \delta^i{}_j\ ,
\eeq
and since only its pure trace part changes, its tracefree part is invariant
\beq
  [\bar L \beta]^i{}_j
  = [L \beta]^i{}_j 
\ .
\eeq
The same is true of the tracefree part of the time derivative term
\beq
 \bar g{}^{ik} \dot {\bar g}_{kj}
     =  g{}^{ik} \dot g_{kj} + 4(\ln\phi)\,\dot{} \, \delta^i{}_j
\ ,
\eeq
which is also invariant, suggesting that the conformal transformation of the tracefree part of the mixed form of the extrinsic curvature should be due entirely to the common factor of the lapse which divides both terms
\beq
  \bar A^i{}_j = \phi^{-6} A^i{}_j\ .
\eeq
This is in fact reinforced by the conformal transformation properties of the divergence operator appearing in the supermomentum constraint, expressed in terms of the pure trace and tracefree parts of the extrinsic curvature.
 
The transformation of the divergence of a symmetric tensor $S^{ij}$ is easily evaluated
\begin{eqnarray}
&&\kern-10pt
\bar S{}^{ij} = \phi^{x-4} S^{ij}  \quad {\rm or}\quad
\bar S{}^i{}_j = \phi^{x} S^i{}_j \rightarrow\\
&&\kern -10pt  
\bar\nabla_j \bar S{}^j{}_i =\phi^x (\nabla_j S^j{}_i +(x+6)S^j{}_i\nabla_j \ln\phi
                           -2 S^j{}_j \nabla_i \ln\phi) \ .
\nonumber
\end{eqnarray}\typeout{!!! explicit kern used here to squeeze displayed equation to left slightly}%
Picking $x=-6$ makes the divergence of a tracefree such tensor also transform by a conformal factor, namely by the weight $-6$ for the covariant (index-lowered) form of the divergence, corresponding exactly to the above transformation due to the reciprocal lapse factor alone.

On the other hand, the trace of the extrinsic curvature is always held fixed \cite{york73}
so the supermomentum constraint
\beq\label{eq:smcon2}
\bar\nabla_j \bar A{}^j{}_i = {\textstyle\frac23} \nabla_i \tau + \bar j_i\ .
\eeq
transforms to
\beq\label{eq:smcon3}
\nabla_j A{}^j{}_i = \phi^6 ({\textstyle\frac23} \nabla_i \tau + \bar j_i)\ .
\eeq
This constraint can be solved by decomposing the tracefree part of the extrinsic curvature into the sum of a transverse traceless part (having zero divergence) and a longitudinal part involving a vector potential $Y$.
However, the covariant form of the divergence of the tracefree Lie derivative operator is invariant (if the vector potential doing the differentiation is invariant)
\beq
 [\bar L Y]^i{}_j =  L Y^i{}_j\ ,
\eeq
while the transverse traceless piece should transform,
so one must include an additional transforming factor in the longitudinal part to get the two pieces to transform consistently. The missing lapse factor corresponding to lapse-corrected derivatives makes the longitudinal part transform correctly by the factor $\phi^{-6}$, so the decomposition transforms unambiguously
\begin{eqnarray}
  \bar A{}^i{}_j &=& \bar A_{(TT)}{}^i{}_j + (2\bar N)^{-1} [\bar L Y]^i{}_j\ ,
\nonumber\\
  A{}^i{}_j &=& A_{(TT)}{}^i{}_j + (2 N)^{-1} [ L Y]^i{}_j\ .
\end{eqnarray}
Imposing the transverse condition on the first term, the vector potential equation then takes the form
\beq
\bar\nabla_j \left((2\bar N)^{-1} [\bar L Y]^j{}_i \right) 
= \bar\nabla_j \bar A{}^j{}_i
\eeq
in terms of the barred variables,
where the right hand side can be replaced using the supermomentum constraint to yield 
\beq
\bar\nabla_j \left((2\bar N)^{-1} [\bar L Y]^j{}_i \right) 
= {\textstyle\frac23} \nabla_i \tau + \bar j_i\ .
\eeq
Alternatively in terms of the unbarred variables, 
\beq
\nabla_j ((2 N)^{-1} [L Y]^j{}_i )
= \phi^6 ({\textstyle\frac23} \nabla_i \tau + \bar j_i)\ .
\eeq
This is the final improved conformal approach introduced by York \cite{py03}. Note that one can rewrite the vector potential equation in the form
\beq
\bar\nabla_j (- \bar A{}^j{}_i + (2\bar N)^{-1} [\bar L Y]^j{}_i ) 
=0\ ,
\eeq
which is the same as the new minimal distortion shift equation for $\beta^i$. Therefore from the previous discussion, the decomposition of the tracefree part of the extrinsic curvature into transverse and longitudinal parts is orthogonal with respect to the full spacetime measure inner product.

To show the relationship between the minimal distortion shift and the vector potential, we must look closer at the initial value problem in terms of the thin sandwich variables.
Given $\bar N$ and $\bar g_{ij}$, starting from any tracefree symmetric tensor $\bar C_{ij}$, one can remove its divergence to get a transverse traceless symmetric tensor 
\begin{eqnarray}
&&   \bar A_{\rm(TT)}^{ij} = \bar C^{ij}-(2\bar N)^{-1} [\bar L V]^{ij}\ ,
\nonumber\\
&&  \bar\nabla_j ((2\bar N)^{-1} [\bar L V]^{ij}) = \bar\nabla_j \bar C^{ij}
\end{eqnarray}
which then determines the transverse traceless part of the barred extrinsic curvature by the conformal rescaling. The longitudinal part $[\bar L Y]^{ij}$ of the tracefree extrinsic curvature is then determined by the supermomentum constraint (\ref{eq:smcon2}), from which the transverse term drops out, leading to 
\beq\label{eq:VP2}
        \bar\nabla_j [ (2\bar N)^{-1} \bar L Y]^j{}_i 
  = \bar\nabla_j \bar A{}^j{}_i
  = {\textstyle\frac23} \nabla_i \tau + \bar j_i\ .
\eeq
The subtracted divergence part can then be combined with the vector potential term
\begin{eqnarray}
  \bar A{}^{ij}
&=& \bar C{}^{ij}-(2\bar N)^{-1} [\bar L V]^{ij} + (2\bar N)^{-1} [\bar L Y]^{ij} 
\nonumber\\
&=& \bar C{}^{ij}+(2\bar N)^{-1} [\bar L (Y-V)]^{ij}
\ ,
\end{eqnarray}
This entire discussion could be transformed to the unbarred variables, or one could start there and transform back. The same equations apply to both sets of variables.

In the thin sandwich picture, comparing this last expression for the tracefree extrinsic curvature with the expression in Eq.~(\ref{eq:Athinsand}), we can make the identifications 
\begin{eqnarray}
&& \bar C_{ij} = -(2\bar N)^{-1} [\dot {\bar g}_{ij}
        - {\textstyle\frac13} \bar g_{ij} \bar g^{kl} \dot g_{kl}]\ ,
\nonumber\\
&& (2\bar N)^{-1} [\bar L (Y-V)]_{ij} = (2\bar N)^{-1}  [\bar L\beta]_{ij}
\ .
\end{eqnarray}
Furthermore, it is then natural to identify the shift with the difference vector field
\beq
   \beta^i = Y^i-V^i\ ,
\eeq
although in general they would only be forced to be equal modulo a conformal Killing vector field of the spatial metric. We recall that on an asymptotically Euclidean slice, conformal Killing vectors do not vanish at infinity and thus they are eliminated.

Suppose we take a spacetime sliced and threaded in zero shift gauge $\beta^i=0$. This implies that the vector potential $Y^i$ at each instant of coordinate time (think of an evolving initial value problem) can be chosen to equal the vector $V^i$ which generates the divergence of the tracefree part of the conformal metric velocity. The remaining transverse traceless piece could then be thought of as corresponding to the time derivative of the evolving dynamical part of the spatial metric.

On the other hand one can choose the vector $V^i$ to be zero, insisting that the conformal metric velocity be transverse, which forces the shift vector field to equal the vector potential, so that it satisfies the vector potential equation and hence
corresponds to a new minimal distortion shift vector field. In this gauge only the ``transverse traceless" part of the conformal metric evolves, i.e., the part of the conformal metric whose (lapse corrected) time derivative is transverse traceless. Thus the gauge condition effectively reduces the evolution of the spatial conformal metric to its ``dynamical" part \cite{york72,york74} which is invariant under spatial diffeomorphisms.

Both possibilities describe the two most useful spatial gauge choices for the Bianchi type IX spacetimes where the spatial diffeomorphism group is exactly the symmetry group $SO(3,R)$ of the rigid body problem used as an analogy \cite{bob80,unified} for understanding the spatial diffeomorphism gauge freedom of generic spacetimes by Fischer and Mardsen \cite{fismar}. The zero-shift gauge corresponds to space-fixed coordinates in the rigid body problem, while the transverse gauge corresponds to body-fixed coordinates, in which the spatial metric is diagonalized exactly as is the moment of inertia tensor in the rigid body problem. Spatial homogeneity makes the analogy much more direct since one does not have to deal with the complication in which one evaluates the spacetime metric before or after a spatial diffeomorphism and only need worry about how the components transform. 
The old minimal distortion shift, due to the incorrect scaling by the lapse function in the old conformal approach to the initial value problem, leads to the shift being proportional to the vector potential, a near miss that seemed a shame several decades ago but that was never pursued \cite{bob80}.

In conclusion, we note that this foregoing discussion elegantly links kinematics and dynamics through proper understanding of the initial value problem together with the new minimal distortion shift vector.

\begin{acknowledgments}
The authors thank Giorgio Ferrarese and Tomaso Ruggeri for organizing the Meeting ``Analysis, Manifolds and Geometric Structures in Physics (Elba, 2004)" in honor of Yvonne Choquet-Bruhat's eightieth birthday, where this work was completed. J.W.Y. thanks the National Science Foundation for support through Grant No.~047762.
\end{acknowledgments}

\end{document}